\newread\testifexists
\def\GetIfExists #1 {\immediate\openin\testifexists=#1
	\ifeof\testifexists\immediate\closein\testifexists\else
        \immediate\closein\testifexists\input #1\fi}
\def\epsffile#1{Figure: #1} 	

\immediate\openin\testifexists=e
	\ifeof\testifexists\immediate\closein\testifexists\else
        \immediate\closein\testifexists\input e\fipsf 
  
\magnification= \magstep1	
\tolerance=1600 
\parskip=5pt 
\baselineskip= 5 true mm \mathsurround=1pt
\hsize=5.3in  
\vsize=7.2in 
\font\smallrm=cmr8

\font\medrm=cmr9  
\font\medit=cmti9

\font\bigbf=cmbx12
 	\def\Bbb#1{\setbox0=\hbox{$\tt #1$}  \copy0\kern-\wd0\kern .1em\copy0} 
	\immediate\openin\testifexists=a
	\ifeof\testifexists\immediate\closein\testifexists\else
        \immediate\closein\testifexists\input a\fimssym.def 
\def\secbreak{\vskip10pt plus .5in \penalty-100\vskip 0pt plus -.4in} 
\def\biggerskip{\vskip6mm plus 2mm}
\def\hugeskip{\vskip9mm plus 3mm}
\def\Narrower{\par\narrower\noindent}	
\def\Endnarrower{\par\leftskip=0pt \rightskip=0pt} 
	\def\ra{\rightarrow}		
          \def\b{\beta}   \def\g{\gamma}  
          \def\D{\Delta}  \def\e{\varepsilon}
              \def\l{\lambda}         \def\L{\Lambda} 
\def\m{\mu}             \def\f{\phi}                
\def\n{\nu}             \def\j{\psi}    
           
            \def\th{\theta}      
                     
\def\w{\omega}  \def\W{\Omega}                          

 \def\LL{{\cal L}}

\def\cl{\centerline}    

\def\ni{\noindent}      \def\pa{\partial}       \def\dd{{\rm d}}        
                 \def\bra{\langle}       \def\ket{\rangle}
 
\def\fn#1{\ifcase\noteno\def\fnchr{*}\or\def\fnchr{\dagger}\or\def
	\fnchr{\ddagger}\or\def\fnchr{\rm\S}\or\def\fnchr{\|}\or\def
	\fnchr{\rm\P}\fi\footnote{$^{\fnchr}$} 
	{\scrunch#1\toe}\ifnum\noteno>3\global\advance\noteno by-5\fi
	\global\advance\noteno by 1}
 	\def\scrunch{\baselineskip=11 pt \medrm}
 	\def\toe{\vphantom{$p_\big($}}
	\newcount\noteno

\def\fract#1#2{{\textstyle{#1\over#2}}}
\def\ffract#1#2{\raise .35 em\hbox{$\scriptstyle#1$}\kern-.25em/
	\kern-.2em\lower .22 em \hbox{$\scriptstyle#2$}}

\def\half{\fract12} \def\quart{\fract14}

\def\part#1#2{{\partial#1\over\partial#2}} 
 \def\ref#1{${\vphantom{)}}^#1$}

\def\bbf#1{\setbox0=\hbox{$#1$} \kern-.025em\copy0\kern-\wd0
        \kern.05em\copy0\kern-\wd0 \kern-.025em\raise.0433em\box0}              
\def\qu{\ {\buildrel {\displaystyle ?} \over =}\ }
\def\ddef{\ {\buildrel{\rm def}\over{=}}\ }
\def\is{\,=\,&}

\def\ref#1{${\,}^{\hbox{\smallrm #1}}$}

\def\Gbar{\raise.13em\hbox{--}\kern-.35em G}
\def\lap{\setbox0=\hbox{$<$}\,\raise .25em\copy0\kern-\wd0\lower.25em\hbox{$\sim$}\,}
\def\glt{\setbox0=\hbox{$>$}\,\raise .25em\copy0\kern-\wd0\lower.25em\hbox{$<$}\,}
\def\gap{\setbox0=\hbox{$>$}\,\raise .25em\copy0\kern-\wd0\lower.25em\hbox{$\sim$}\,}
   \def\newsect#1{\secbreak\noindent{\bf #1}\medskip}
\def\inv{^{\rm inv}}
\def\Bar#1{\overline{#1}}
\def\Tr{{\rm Tr\,}}
 \rightline{SPIN-1998/19}
 \rightline{hep-th/9812204} 
 \medskip
 
\cl{\bigbf TOPOLOGICAL ASPECTS OF}\smallskip
\cl{\bigbf QUANTUM CHROMODYNAMICS}\medskip
\cl {Erice lecture notes, August/September 1998}
\hugeskip

\cl{\bf Gerard 't Hooft }
\biggerskip
\cl{Institute for Theoretical Physics}
\cl{University of Utrecht, Princetonplein 5}
\cl{3584 CC Utrecht, the Netherlands}
\smallskip
\cl{and}
\smallskip
\cl{Spinoza Institute}
\cl{Postbox 80.195}
\cl{3508 TD Utrecht, the Netherlands}
\smallskip\cl{e-mail: \tt g.thooft@phys.uu.nl}
\cl{internet: \tt http://www.phys.uu.nl/\~{}thooft/	}
\hugeskip
\ni{\bf Abstract:}\Narrower
 Absolute confinement of its color charges is a natural property of
 gauge theories such as quantum chromodynamics. On the one hand, it
 can be attributed to the existence of color-magnetic monopoles, a
 topological feature of the theory, but one can also maintain that
 all non-Abelian gauge theories confine. It is illustrated how 
 ``confinement'' works in the $SU(2)$ sector of the Standard Model,
 and why for example the electron and its neutrino can be viewed  
 as $SU(2)$-hadronic bound states rather than a gauge doublet.
 The mechanism called `Abelian projection' then puts the Abelian
 sector of any gauge theory on a separate footing.
\Endnarrower
\bigskip\newsect{1.  RUNNING COUPLING STRENGTHS.}

The Lagrangian of an arbitrary renormalizable gauge theory in general
takes the form
$$\eqalign{\LL\inv=-\quart G_{\m\n}G_{\m\n}-\half(D_\m\f)^2-\Bar\j\g D\j\cr
-V(\f)-\Bar\j\big(S(\f)+i\g_5P(\f)\big)\j\,,}\eqno(1.1)$$
where, among the usual definitions, the covariant derivatives $D_\m$
are defined using representation matrices $T^a$ and $U^a$ associated to
the gauge generators $\L^a$ of the gauge group, as follows:
$$D_\m\f\ddef\pa_\m\f+T^aA_\m^a\f\quad,\qquad D_\m\j\ddef\pa_\m\j+U^aA_\m^a\j\ .\eqno(1.2)$$

The complete set of beta functions for the theory is then described by the
algebraic expression\ref1
$${\m\dd\over\dd\m}\LL\inv={1\over8\pi^2}\D\LL\inv\,,\eqno(1.3)$$
and after a long calculation\ref2, one finds that
$$\eqalign{\D\LL\inv\is -\quart G_{\m\n}^aG_{\m\n}^b\Big(\fract{11}3C_1^{ab}
-\fract23C_2^{ab}-\fract16C_3^{ab}\Big)\cr
	&-\,\D V(\f)-\Bar\j(\D S+i\g_5\D P)\j\,;}\eqno(1.4)$$
$$\eqalign{\D V\is\quart(\pa_i\pa_j V)^2+\fract32\pa_i V(T^2\f)_i+\fract34(\f T^aT^b\f)^2\cr
	&+\f_iV_j\Tr(S_iS_j+P_iP_j)-\Tr(S^2+P^2)+\Tr\big[S,P\big]^2\,;}\eqno(1.5)$$
$$\eqalign{\D W\is\quart W_iW_i^*W+\quart WW_i^*W_i+W_iW^*W_i\cr
	&+\fract32U_R^2W+\fract32WU_L^2+W_i\f_j\Tr(S_iS_j+P_iP_j)\,;}\eqno(1.6)$$
$$W\ddef S+iP\quad;\qquad U\ddef\half(1+\g_5)U_L+\half(1-\g_5)U_R\,.\eqno(1.7)$$
 \midinsert\cl{\epsffile{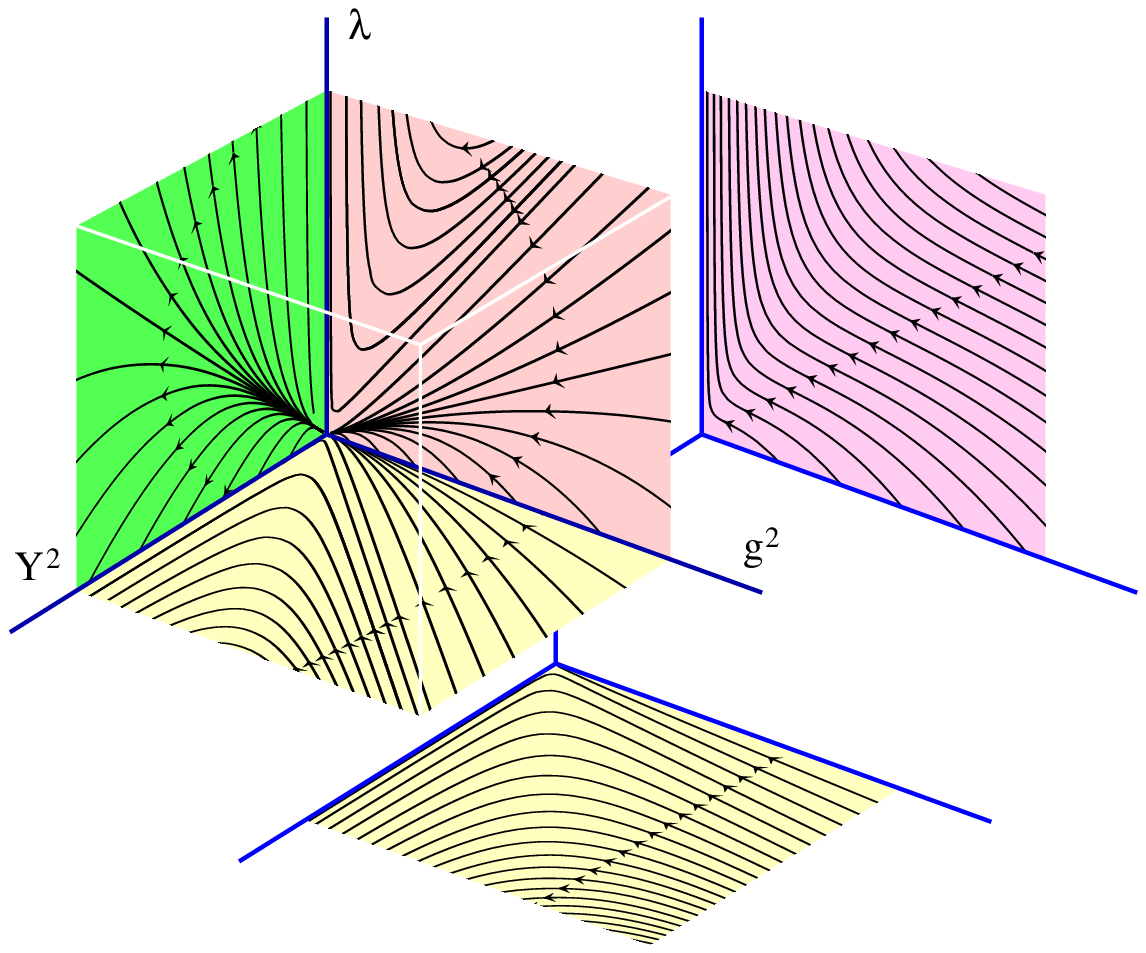}}
 \Narrower Fig.~1. Three dimensional plot of the running couplings in a 
  general theory. The flow lines are projected against the three planes at the
  back of the cube. Alternative possibilities are shown as well. \Endnarrower\endinsert	
\ni Here, a short-hand notation was used: $V_i=\pa_iV=\pa V/\pa\f_i$. 
These expressions can be derived from diagrammatic calculations in combination
with arguments to exploit local gauge invariance. One finds that, quite generally,
the signs implied for the beta functions are universally fixed. If $g$ stands for
the collection of gauge coupling constants, $Y$ for the Yukawa couplings and
$\l$ for the scalar self interactions, we can write in short-hand:
$$\eqalignno{{\m\dd\over\dd\m}g\ddef&\b_g(g)\ =(-R_1+R_2N_f+R_3N_s)g^3\,;&(1.8)\cr
{\m\dd\over\dd\m}Y\ddef&\b_Y(Y)=R_4Y^3-R_5 g^2Y\ ;&(1.9)\cr
{\m\dd\over\dd\m}\l\ddef&\b_\l(\l)\ =R_6\l^2-R_7g^2\l+R_8g^4+R_9y^2\l-R_{10}Y^4\,.
&(1.10)}$$
Here, $R_1,\,\dots,\,R_{10}$ each may actually represent matrices of coefficients,
but quite generally they are positive (with the exception for $R_1$ in the Abelian case).
However, the relative magnitudes of these coefficients may vary considerably, and
henceforth, the resulting renormalization flow patterns generated may fall in different
possible classes. In general, the result is as depicted in Fig.~1. The arrows point into
the ultraviolet regime.

The infrared behaviour of a theory such as QCD is strongly determined by these
beta functions. In most cases, a non-Abelian theory will have a negative beta function
for the gauge coupling constant; only if there are many massless spinor and/or scalar
particles around, might one encounter the case that this beta function is positive or
zero, such as in the $N=4$ super symmetric theory.\ref3 The spinors and scalars must also
be protected against developing mass terms, and super symmetry can do exactly this.
We must also be aware that matter fields in representations higher than the adjoint
one, in general have large Casimir coefficients $R_2$, $R_3$, so they, also, tend to
flip the sign of the beta function.

If beta is positive, or zero, the theory can stay in the perturbative phase in the entire
infrared domain, and its infrared behaviour reminds one of that of a plasma. Otherwise,
strong interactions force the system to condense in some way or other. There are then
three basic possibilities:\ref4
\item{1.} A diagonal ({\it i.e.} Abelian) subgroup of the gauge group survives
undisturbed. Long-range Coulomb forces will then dominate over large distances, as in
quantum electrodynamics (QED). Charged particles usually will not be protected against
mass generation (exotic exceptions can be visualized), so that the Maxwell system will
be scale invariant. However, in contrast to QED, there will also be isolated magnetic charges,
usually also with masses. In the far infrared, therefore, the pure Maxwell fields are the
only long-range fields. Since magnetic charges and electric charges both occur, this
condensation mode may be called `self-dual'.
\item{2.} There is a complete Higgs mechanism. No long-range electric fields survive.
Thus also no monopoles survive in the infra-red. All forces are short-range, being
mediated by massive heavy gauge bosons. This situation can be conveniently described
in theories with scalars, using perturbation expansion.
\item{3.} Complete Higgs mechanism in the `magnetic sector'. This condensation mode is dual to
the previous case. All particle species that entered in the ultraviolet description as
non-trivial representations of the gauge group (`quarks'), will be confined into colorless
(`hadronic') bound states, being bound by vortex-like field configurations. This is
what we call confinement.
\newsect{2. BETWEEN HIGGS AND CONFINEMENT.}
Both in the Higgs phase (case 2) and in the confinement phase (case 3), there is what
we call a {\it mass gap}. The lightest existing physical particle in the theory has a 
non-vanishing mass $m_0$, and therefore there are no physical states in Hilbert space with
energy between that of the vacuum ($H|\emptyset\ket=0$) and the one particle state
($H|\,p_1=0\,\ket=m_0c^2|\,p_1=0\,\ket$). The distinction between these two phases is therefore not in
the energy spectrum alone, but rather in the quantum numbers of the existing states.
However, this distinction is not always clear-cut. Indeed, if the Higgs field is
{\it in the fundamental representation of the gauge group}, there is {\it no formal
distinction at all\/} between the Higgs phase and the confinement phase, a situation
comparable to what distinction we have between the gaseous phase and the liquid phase
of water. Following a curve in the pressure--temperature plot above the critical point,
we see that a continuous transition between the one and the other is possible. We shall
now illustrate this for the standard weak interaction theory, where the gauge group is
$SU(2)\otimes U(1)$, broken by a Higgs in the fundamental representation of $SU(2)$.

To make our point, let us consider the following ``QCD-inspired model" for the electro-weak
force.\ref5 The gauge group is $SU(2)\otimes U(1)$, but we concentrate on the $SU(2)$ forces,
while treating $U(1)$ as a fairly insignificant perturbation. There are leptons and quarks.
The latter also have color $SU(3)$ quantum numbers, which will be the usual ones, but we
do not consider the effects of the strong force. The leptons and the quarks come in exactly
the same representations of the gauge groups as in the usual theory. The difference between
the model described below and the standard model is only in our description of the physical
particles. {\it All physical particles are singlets under $SU(2)$}, because there is
`weak-color'\fn{Awaiting a better terminology, we use the word `weak-color' to indicate the 
weak {\medit SU\/}(2)  quantum numbers.} confinement! 

The $U(1)$ gauge boson will be identified with the photon. Although it will mix somewhat
with the $Z^0$, the latter will not be regarded as a gauge boson, so we do not say that
$U(1)$ and $SU(2)$ are mixed.

The $SU(2)$ weak-gluons are not at all identified with the $W^\pm$ or the $Z^0$. They will
be as far from physical as the strong color gluons are.

\settabs 9\columns
The fundamental $SU(2)$ doublet fields are also unphysical in the sense that they are treated
as the weak versions of quarks; let us call them weak-quarks.  We have the leptonic weak-quarks:
\+&& $\ell_i$&(spin $\half$, left handed),\cr\ni 
we have quark weak-quarks:
\+&& $q_{i}$&(spin $\half$, left handed, $SU(3)^{\rm color}$ triplets),\cr\ni
and Higgs weak-quarks:
\+&& $h_i$&(spin 0).\cr\ni
\ni For simplicity, we consider a single generation of leptons and quarks. All these fields
have the usual assignment of $U(1)$ charges. Of course, the $U(1)$ transformations commute with
the $SU(2)$ transformations, so that the $U(1)$ charges do not depend on the weak-color
index $i$. In addition, there will be right handed spinor fields being $SU(2)$ singlets.
They are just as in the standard model.

All  physical particles are weak-hadrons. In principle, one may distinguish weak-baryons
(built out of two weak-quarks), and weak-mesons (built out of a weak-quark and a weak-antiquark),
but, since the weak-color group is $SU(2)$, a weak-quark field $\j_i$ is linearly
equivalent to a weak-antiquark field, $\j^i =\e^{ij}\j_j$, so that the distinction
between weak-baryons and weak-mesons depends on our conventions in defining the elementary
fields. The convention chosen by the author will become apparent shortly.

The weak-hadrons built out of pairs of quark and lepton fields will exist, but they are highly
unstable, and not of our primary concern. Much more interesting will be the hadrons that
contain one or more of the scalar weak-quarks $h$ and/or $\bar h$. Important weak-mesons are:
\item{--}  The bound state $\bar h\,\ell$, to be identified with the {\it neutrino}.
\item{--}  The bound state $\bar h\,q$, to be identified with the (left handed) up-quark.
\item{--}  The bound state $\bar h\,h$ without orbital angular momentum ($S$ state); it is the
physical Higgs particle.
\item{--}  If its orbital angular momentum is 1 ($P$ state), it is the $Z^0$.
\par\ni The important weak-baryons are:
\item{--}  The bound state $h\,\ell$, being the (left handed) electron;
\item{--}  The bound state $h\,q$, being the (left handed) down quark.
\item{--}  The $P$ state $h\,h$ being the $W^-$ (there is no $S$ state).
\par\ni The logic of these assignments will become evident shortly. The fact that all physical
particles are weak-color singlets is to be regarded exactly as in QCD. There is, however,
one important distinction. In this theory, one can do accurate computations, because
we have at our disposal a {\it reliable perturbative technique.}

Since we have a scalar field in the elementary representation, we can use it
to fix the gauge. Let us consider the unitary gauge obtained by rotating this scalar
field in the up-direction. We write
$$h_i=\pmatrix{F+h_{(1)}\cr 0\hfill}\,,\eqno(2.1)$$
Here, $F$ is a suitably chosen number, and $h_{(1)}$ is a residual field. The scalar 
self-interaction now happens to be such that the number $F$ is large, and the $1/F$
expansion can be made. Note, however, that $F$ can {\it always\/} be chosen to be 
non-vanishing, as soon as the scalar field fluctuates, so that our opportunity to
perform this expansion does not require a double-lobed potential in the Higgs
self-interaction. A double-lobed potential is required only if we insist $F$ to be so large that
the perturbative expansion converges rapidly.

With this choice of gauge, the above weak-mesonic composite fields can be written
as
$$\eqalignno{\bar h\,\ell\is \Big(F+h_{(1)},\ 0\,\Big)\cdot
\pmatrix{\ell_1\cr\ell_2}=F\,\ell_1+\dots\,,&(2.2)\cr
\bar h\,q\is F\,q_1+\dots\,,&(2.3)\cr
\bar h\,h\is F^2+2F\,h_{(1)}+\dots\,,&(2.4)}$$
and the weak-baryonic fields as
$$\eqalignno{\e^{ij}h_i\,\ell_j\is F\,\ell_2+\dots\,,&(2.5)\cr
\e^{ij} h_i\,q_j\is F\,q_2+\dots\,.&(2.6)\cr}$$
The dots here stand for small higher order corrections.
The $P$-states contain a derivative of the fields:
$$\eqalignno{\bar h\,D_\m h\is  F(-\half i gW^3_\m )F+\dots\ =\ -\half i gF^2\,W^3_\m\,,
&(2.7)\cr
\e^{ij}h_iD_\m h_j\is F(-\half ig)(W^1_\m+i W^2_\m)F+\dots\ =\ -\half i gF^2\,W^-_\m\,.
&(2.8)}$$

We observe that, {\it within this gauge choice}, and apart from insignificant
numerical coefficients, we identified the left-handed fermion fields as
$$\pmatrix{\ell_1\cr \ell_2}=\pmatrix{\n\cr e}\ ,\qquad\pmatrix {q_1\cr q_2}=
\pmatrix{u\cr d}\ ,\eqno(2.9)$$
and the $SU(2)$ vector fields coincide with the usual definitions as well.
From here on, all calculations for the electro-weak theory are performed exactly in
the way they are usually done. The ``confining" model described above is mathematically
identical to the Standard Model.

The important conclusion from this section is that the fundamental difference
between the electro-weak theory and QCD is {\it not\/} that the electroweak
gauge fields fail to confine their charges, since the electroweak charges can
be said to be confined exactly as in QCD. The fundamental difference is that,
in contrast to QCD, the electro-weak theory admits a very good perturbative
approximation technique -- standard perturbation theory. In fact, in all
cases that we have such a perturbative technique at our disposal, we may fix the
gauge anyway we like. We could use the `unitary gauge', as described above,
or, alternatively, one of the many possible renormalizable gauges. The price
one then pays is the emergence of Faddeev-Popov ghosts. In standard perturbation
theory, we have learned how to disentangle the physical states from the ghosts,\ref6
which makes the procedure acceptable anyway. But, if we
do not have such a perturbative approximation technique, gauge-fixing must be
carried out much more judiciously. In a non-perturbative system, such as QCD,
it is often to be preferred to avoid the emergence of ghosts due to gauge-fixing,
in particular if we wish to identify exactly what are the physical states.
\newsect{3. GAUGE-FIXING.}
The physical reason why ghosts may show up, is the non-local nature of the
gauge-fixing procedure. If we demand, for instance,
$$\pa_\m A_\m=0\,,\eqno(3.1)$$
then the transition from some other gauge choice to this one requires knowledge
of the field values of a given configuration over all of space-time. Since gauge
transformations do not affect physical information, the information transmitted over
space-time in order to realize the gauge (3.1), is unphysical. This is the 
explanation of the emergence of ghosts. 

We can avoid ghosts, if the gauge fixing at any point $x$ in space-time, is done in 
such a way that no knowledge of the field values in points other than the point $x$
is needed.

\def\Im{{\rm Im}}\def\Re{{\rm Re}}
The simplest example is the gauge group $SU(2)$. The gauge-fixing can be carried out
at any point $x$ of space-time, without reference to other points, if we have to
our disposal a scalar field transforming as a fundamental representation. The gauge
choice
$$\pmatrix{\f_1\cr\f_2}=\pmatrix{F+h_{(1)}\cr 0\hfill}\,,\eqno(3.2)$$
where $h_{(1)}$ is a real field and $F$ an arbitrary normalization constant, fixes
the gauge completely: the conditions $$\Im(\f_1)=0\ ;\qquad\f_2=0\,,\eqno(3.3)$$ form three
constraints for the three gauge generators $\L^a$. There is no gauge ambiguity left.
{\it One should not interpret $F$ is a ``vacuum expectation
value"}. Statements such as $$\bra\f\ket\qu \pmatrix{F\cr 0}\ne 0\,,\eqno(3.4)$$ have no
physical meaning, just because $\f$ is not gauge-invariant.

Notice that, in perturbation theory, a pure $SU(2)$ theory with scalars in the fundamental
representation exhibits a {\it global\/} $SU(2)$ symmetry in addition to the
local one. This is because the self-interaction,
$$V(\f)=\quart\l|\f|^4\pm\m|\f|^2\,,\eqno(3.5)$$
only depends on the combination $ \Re(\f_1)^2+\Im(\f_1)^2+\Re(\f_2)^2+\Im(\f_2)^2$,
which has the symmetry group $SO(4)=SU(2)^{\rm local}\otimes SU(2)^{\rm global}$.
The gauge choice (3.2) connects the local $SU(2)$ to the global one, and this is why
the local doublets, after gauge-fixing, combine into doublets with a global
$SU(2)$ symmetry. It is only the global $SU(2)$ symmetry that is reflected in the
hadronic mass spectrum. The local and the global $SU(2)$ symmetries are often confused.

The global $SU(2)$ symmetry continues to be exact beyond perturbation expansion,
simply because the global group is spanned by the two complex fields $\f_i$ and
$\f'_i=\e_{ij}\f^{*j}$. However, these objects carry different $U(1)$ charges, and
this is why $U(1)$ breaks this symmetry, so that the $U(1)$ force generates
mass differences within members of doublets, as well as mass splittings between
$W^\pm$ and $Z^0$.

{\it Topological features\/} are present when we use scalar elementary
representations for gauge-fixing. They occur whenever, accidently, all four
field components vanish. At isolated points in space-time, one may have
$$\f({\bf x},t)=0\,,\eqno(3.6)$$
and this is where instantons will occur. They break the chiral symmetries
through the anomalies, as usual.
\newsect{4. THE ABELIAN PROJECTION.\ref4}

How do we fix the gauge, while avoiding ghosts, if there are no scalars in the elementary
representation? We try to do the next-best thing: take any bosonic field, or
combination of bosonic fields, that transform locally (i.e., without derivatives $\pa_\m\L$)
under a local gauge transformation, to find some preference frame in gauge space.
For instance, we could use the covariant magnetic curl field in the $z$ direction, $G_{12}$.
It transforms under infinitesimal gauge transformations as
$$G_{12}\ra G_{12}'=G_{12}+[\L,\,G_{12}]\,.\eqno(4.1)$$
The important point is, that this is not the fundamental, but the adjoint representation
of the gauge group, and there are elements of the gauge group under which these fields
do not transform at all (the `center' of the gauge group). This is why a new situation
arises when we do this. Quite generally, when there are no extra scalar fields around,
and we can only use (combinations of) the gauge fields themselves, then we have only
those representations that are invariant under the center (the `non-exotic' representations
of $SU(3)$, for instance). Let us take the example of the adjoint representation, and
take the field $G_{12}$, just to be specific.

$G_{12}$ does not fix the gauge completely; all we can do with it is, constrain ourselves to
the gauge choice such that this field is diagonalized:
$$G_{12}\ra\pmatrix{*\ &&0&\cr &*\ &&\cr 0&&*\ &\cr &&&\cdot.\ }\,.\eqno(4.2)$$

This then leaves unfixed a subgroup of the local gauge group. The group of transformations
that do not affect the form (4.2) of this field, are the set that commute with $G_{12}$.
In general, this group, called the {\it Cartan subgroup}, has the form $U(1)^{N-1}$, if the original group was $SU(N)$,
the $-1$ coming from the restriction that the determinant must stay equal to one.
A separate procedure is needed to remove this remaining gauge redundancy, but we may
observe, that the Cartan subgroup is an Abelian gauge group, and so, our theory at this stage is
just Maxwell's theory; we can fix the gauge any way we like, because quantum electrodynamics
tells us exactly how to compute things here. Our theory has been reduced to an ordinary
Abelian system.

There is, however, a topological novelty, not shared by electrodynamics: the field $G_{12}$
employed to fix the gauge may have coinciding eigenvalues at certain points in space-time.
If we investigate the nature of those subspaces of space-time where this might happen,
we find that the condition $\l_{(1)}=\l_{(2)}$, for two adjacent eigenvalues $\l$, gives us 3
constraints (the number of constraints needed to turn an arbitrary $2\times2$ hermitean
matrix into a multiple of the identity matrix). Therefore, we expect these points to be
particle-like. Indeed, one can easily verify that these particle-like objects are
{\it magnetic monopoles\/} with respect to the $U(1)$ photons. This is actually a
familiar situation. Taking, instead of $G_{12}$, some additional scalar field (in the adjoint
representation), then what we have here is a Higgs theory where $SU(2)$ `breaks down into'
$U(1)$, and the zeros of the scalar field generate the magnetic monopole charges.

The procedure described here is known as `the Abelian projection'. It is a partial ghost-free
gauge fixing procedure that turns any non-Abelian gauge theory into a theory with only Abelian
gauge fields, electric {\it and magnetic\/} charges. In order to understand the long-distance
features of this theory, one must investigate the ways in which these various types of charges
may Bose-condense. If the magnetic charges Bose-condense, we have permanent quark confinement.

What happens if we use some other field combination to fix the gauge? Suppose that all we have is the 
gauge fields themselves. Let $Z$ be an element of the {\it center\/} of the gauge group:
$$[Z,\W(x)]=0\ ,\qquad \forall\,\W\,.\eqno(4.3)$$
If the gauge group is $SU(N)$, these elements form the subgroup ${\Bbb Z}(N)$, the set of
matrices of the form $e^{2\pi ik/N}\,\Bbb I$, where $k$ is an integer. If we had at our disposal
a scalar field in the elementary representation, using it to fix the gauge would also remove
the freedom to perform center gauge transformations. But the gauge vector fields $A_\m(x)$
are invariant under center element transformations, and therefore they cannot be used
to remove the center redundancy. 

The center of the gauge group gives rise to a new topological object when we use a center-invariant
gauge fixing procedure: a vortex. Consider an operator in Hilbert space defined starting from a
closed curve $C$. Following another closed curve $C'$ interlooping once with $C$, we can
consider a gauge transformation $\W$ which varies from $\Bbb I$ to a non-trivial center element
$Z$ while we loop around the curve $C'$. The element $Z$ can only depend on the looping index.
The gauge field $A_\m$ will be continuous, except on the curve $C$ itself, where a singularity 
develops that has to be smeared a bit. This defines a (very slightly smeared) vortex on $C$:
a magnetic vortex, labled by the index $k$ of the center element $Z$.  

The existence of this magnetic vortex implies the absence of certain magnetic charges
that could have been attached to its end points. The Abelian theory we now get
may still have magnetic charges, but the magnetic charges present will not saturate the
Dirac condition $Q\cdot g_m =2\pi k$ (where $Q$ is the unit of electric charge, and $g_m$
the smallest existing magnetic charge): not all values of $k$ will occur, usually, $k$ will
only be a multiple of $N$. The magnetic vortex described above gets an electric counterpart
if we allow the magnetic charges from the Abelian projection to condense. It is this
electric vortex that binds quarks inside hadrons. In short: in the Higgs phase, the magnetic
vortex is stable, in the confinement phase, the electric vortex is stable.

What happens to this vortex if we make the transition to a scalar field in the elementary
representation to fix the gauge? In that case, the vacuum is saturated with condensed
particles with the fundamental gauge charge. They provide new end points to the electric vortices,
and, since they are abundant, this completely destabilizes the electric vortex, and the
confinement mechanism is made invisible. This is why, in the electro-weak theory, we
usually do not consider the leptons and quarks as being confined. Strictly speaking, they
still form weak-hadrons, by attaching themselves to Higgs particles, as it was described
in Sect.~2, but we can just as well employ the standard description.
\newsect{5. THE EFFECTS OF INSTANTONS ON CONFINEMENT.}
Instantons are more fully explained in other lectures, see Refs\ref{7, 8}.
Here, we give a brief summary.

Instantons occur whenever there is an $SU(2)$ subgroup in the gauge group, i.e.,
in all non-Abelian gauge theories. Consider the temporal gauge,
$$A_0=0\,,\eqno(5.1)$$
In this gauge, the Yang-Mills Lagrangian reads
$$\LL^{\rm YM}=\half(\pa_0 A_i)^2-\quart G_{ij}G_{ij}=\half{\bf E}^2-\half{\bf B}^2\,.
\eqno(5.2)$$
The temporal gauge leads to a description invariant under time-independent
gauge transformations. Therefore, the Hamiltonian generated by this Lagrangian
will commute with gauge transformation operators $\W$:
$$[H,\,\W]=0\,.\eqno(5.3)$$
This means that all states in the Hilbert space associated to this Lagrangian,
can be chosen to be eigenstates of $\W$:
$$\W|\j\ket=\w|\j\ket\,.\eqno(5.4)$$
Now, since $\w$ will be time-independent, whereas $\W$ may well be space-dependent,
this would violate Lorentz invariance unless $\w=1$. Therefore, it is advised to limit ourselves
to the subspace of Hilbert space with $\w=1$, which is the trivial representation 
of the gauge group. 

There is one exception, however. While keeping $\w=1$ for all gauge transformations
that are continuously connected to the identity, we may make an exception for
{\it topologically non-trivial\/} gauge transformations. Since the space of $SU(2)$
transformations form an $S_3$ sphere, we can consider the transformation obtained
by making a topologically non-trivial mapping of this sphere onto 3-space. The class
of these transformations form a discrete set, labled by the winding index $k\in \Bbb Z$.

Gauge transformations should be unitary in Hilbert space, therefore, $|\w|=1$.
Furthermore, since these topological gauge transformations form a group, the states
must be a representation of this group. We find that, if $\W_N=(\W_1)^N$ is the transformation
with winding number $N$, the eigenvalue equation must take the form
$$\W_N|\j\ket=e^{iN\th}|\j\ket\,,\eqno(5.5)$$
where $\th$ can be any angle between 0 and $2\pi$. In fact, $\th$ is a new, conserved
constant of the theory, to be added to the coupling constant $g$ as a fixed parameter
describing the dynamics. It is called the instanton angle. Instantons are transitions
of the gauge system into topologically rotated configurations. In general, these are tunnelling
events.

The angle $\th$ can be incorporated in the Lagrangian by writing
$$\LL^{\rm inv}=-\quart G_{\m\n}G_{\m\n}+{\th i g^2\over 32\pi^2}G_{\m\n}\tilde G_{\m\n}
+A_\m J_\m\,.\eqno(5.6)$$

From this, one can derive that a magnetic monopole will carry not only a
magnetic charge $g_m$, but also\ref9 a fractional electric charge $Q$,
$$Q={\th g^2\over4\pi^2} g_m\,;\qquad g_m={N\over2\pi g}\,,\qquad\hbox{$N$ is integer.}\eqno(5.7)$$
In addition, the monopole may carry integral units of electric charge, but these can
freely vary, as charged particles are picked up in the environment.

Varying $\th$ continuously from 0 to $2\pi$, we notice that the electric charge of
a monopole varies from 0 to one unit. Now, for the confinement mechanism, it is
very important to know the electric charge of the condensed objects. Adding an
electric charge to the condensed monopoles would lead to different characteristics
of the confinement mechanism. From this, one can deduce that as $\th$ is continuously
varied from 0 to $2\pi$, a phase transition must occur somewhere along the path.
Naturally, one may assume that this happens exactly half-way, that is, at $\th=\pi$,
but also more exotic confinement phases may be imagined, such that there are several
phase transitions, not just one. The in-between phases are referred to as
`oblique confinement'. A fuller account of oblique confinement is to be found
in Refs\ref{4, 10}.

\newsect{REFERENCES}

\item{1.}G.~'t Hooft, {\it Nucl.~Phys. \bf B61} (1973) 455.
\item{2.}G.~'t Hooft, {\it Nucl.~Phys. \bf B62} (1973) 444.
\item{3.}S.~Ferrara and B.~Zumino, {\it Nuc.~Phys. \bf B79} (1974) 413;
D.R.T.~Jones, {\it Phys.~Lett. \bf 72B}(1977) 199; S.~Mandelstam, 
{\it Nucl.~Phys. \bf B213} (1983) 149.
\item{4.}G.~'t Hooft, Acta Phys. Austr., Suppl.{\bf 22} (1980) 531;
 {\it Nucl.~Phys.} {\bf B190} (1981) 455;  
{\it Phys.~Scripta \bf 25} (1982) 133.
\item{5.}T.~Banks and E.~Rabinovici, {\it Nucl.~Phys. \bf B160} (1979) 347;
     G.~'t Hooft, in  "Recent Developments in Gauge Theories", Carg\`ese  1979, 
     ed. G.~'t Hooft et al., Plenum Press, New York, 1980, 
     Lecture II, p.~117.
\item{6.}G.~'t Hooft, {\it Nucl.~Phys.} {\bf B33} (1971) 173;
 {\it Nucl.~Phys.} {\bf B35} (1971) 167; G.~'t Hooft and M.~Veltman, {\it Nucl.~Phys. \bf B50} (1972) 318 
\item{7.}S.~Coleman, in {\it The Whys of Subnuclear Physics}, Erice proceedings 1977,
Ed. A.~Zichichi, Plenum Press, New York 1979, p.~805.
\item{8.}G. 't Hooft, {\it Phys.~Repts. \bf 142} (1986) 357.
\item{9.}E. Witten, {\it Phys.~Lett. \bf B86} (1979) 283. 
\item{10.}E.~Rabinovici and J.~Cardy, {\it Nucl.~Phys. \bf 205} (1982) 1; 
G.~'t Hooft, in {\it Confinement, Duality and Nonperturbative Aspects of QCD},
P.~van Baal, Ed., NATO ASI Series, 1998 Plenum Press, New York and London, p. 379.

\bye